\documentclass[twocolumn,showpacs,preprintnumbers,amsmath,amssymb]{revtex4}

\usepackage{graphicx}
\usepackage{dcolumn}
\usepackage{bm}
\begin{document}
\title{ Anisotropic inelastic scattering and its interplay with superconductivity in URu$_{2}$Si$_{2}$}
\
{\author{Zengwei Zhu$^{1,2}$, Elena Hassinger$^{3}$, Zhu'an Xu$^{2}$, Dai Aoki$^{3}$, Jacques Flouquet$^{3}$ and Kamran Behnia$^{1}$}
\affiliation{ (1) Laboratoire Photons Et Mati\`ere (UPMC-CNRS), ESPCI, 75005 Paris, France\\
(2) Department of Physics, Zhejiang University, Hangzhou 310027, China\\
(3)DRFMC/SPSMS,  Commissariat \`a l'Energie
Atomique, F-38042 Grenoble, France}
\date { July 2, 2009 }

\begin{abstract}
In contrast to almost all anisotropic superconductors, the upper critical field of URu$_{2}$Si$_{2}$ is larger when the field is oriented along the less conducting direction. We present a study of resistivity and Seebeck coefficient extended down to sub-Kelvin temperature range uncovering a singular case of anisotropy. When the current is injected along the c-axis URu$_{2}$Si$_{2}$ behaves as a low-density Fermi liquid. When it flows along the a-axis, even in presence of a large field, resistivity remains T-linear down to T$_{c}$ and the Seebeck coefficient undergoes a sign change at very low temperatures. We conclude that the characteristic energy scale is anisotropic and vanishingly small in the basal plane.
\end{abstract}


\maketitle

\begin{figure}
\resizebox{!}{0.7\textwidth}
{\includegraphics{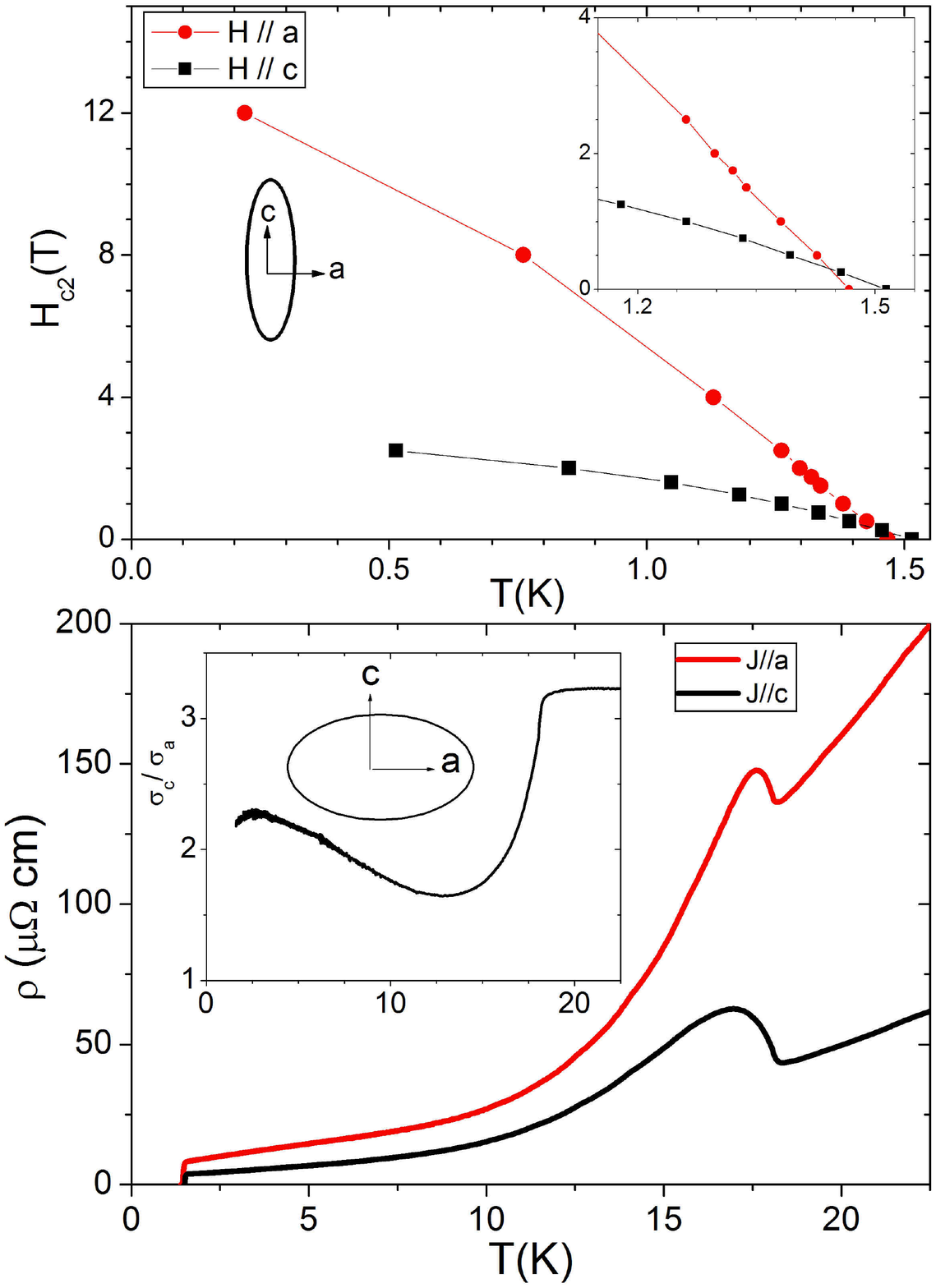}}\caption{ a) Upper critical field in URu$_{2}$Si$_{2}$ for two orientations of the magnetic field. Values correspond to the mid-point resistive transition. The inset is a zoom close to T$_{c}$ highlighting the threefold difference in the slopes close to T$_{c}$. The two sample studied in the two configurations present a slight difference in their zero-field T$_c$s. b) Zero-field resistivity of the same samples. Inset shows the anisotropy of the conductivity. Ellipsoids schematically represent the anisotropy of the Fermi surface according to each probe.}
\end{figure}

The enigma of the phase transition at T$_O$=17 K in URu$_2$Si$_2$ continues to inspire numerous investigations\cite{santander,aoki,shishido,elgazzar,levallois,villaume,hassinger,oh}. The identity of the order parameter emerging below this temperature remains elusive. The radical reconstruction of the Fermi surface accompanying this phase transition and leading to a drastic drop in carrier concentration is now well documented\cite{maple, bel,behnia}. It seems to persist when the ground state switches from HO to antiferromagnetic (AF) state at $P_{\rm x}\simeq 0.5\,{\rm GPa}$\cite{hassinger}, suggesting an identical wave vector for the two states, $\mbox{\boldmath $Q$}_{\rm HO}=\mbox{\boldmath $Q$}_{\rm AF}=(0,0,1)$~\cite{aoki,villaume}. The exotic superconductivity of the surviving dilute heavy electrons\cite{kasahara} disappears at $P_{\rm x}$ as well as the resonance detected in inelastic neutron scattering experiments $E_0\sim 2\,{\rm meV}$ for $\mbox{\boldmath $Q_0$}=(1,0,0)$}.

It has been known for a long time that URu$_2$Si$_2$ is an anisotropic electronic system.  Magnetic susceptibility is 4 to 5 times larger when the field is along the c-axis\cite{palstra}. Charge conduction along the c-axis is 2 to 3 times lower than in the basal plane\cite{palstra2}. The
upper critical field is 3 to five times larger when the field is oriented in the basal plane\cite{brison}. Finally, the  Seebeck coefficient is roughly twice larger when the current flows along the c-axis\cite{sakurai}. Of the three bands detected by de Haas-van Alphen studies, the larger one is isotropic, while the two smaller ones show a modest anisotropy less than two\cite{ohkuni}. Until now, these experimental facts have never been all discussed together.

In this paper, we present a study of two transport coefficients, namely resistivity and Seebeck coefficient, at temperatures well below the superconducting critical temperature. We found that when the current flows along the c-axis, the resistivity displays a T$^2$ behavior and the Seebeck coefficient is linear in temperature. Moreover, and as expected for such a low-density system, the relevant prefactors are enhanced by an order of magnitude. On the other hand, when the current flows along the a-axis, the normal-state resistivity remains linear down to the lowest temperature explored and the Seebeck coefficient displays a non-trivial temperature dependence with a sign change at low temperatures. Thus, in URu$_2$Si$_2$, we encounter a singular case of anisotropic inelastic scattering. For a current flowing along the c-axis, the characteristic energy scale is large enough to find a Fermi-liquid behavior in our temperature range of investigation.  In the basal plane, On the other hand, this energy scale is small enough to impede the emergence of well-defined Landau quasi-particles down to the lowest explored temperatures (0.15 K). Interestingly, the anisotropy of superconductivity is such that the spatial extension of the Cooper pairs is anomalously enhanced in the basal plane where the electron-electron scattering is stronger.

 \begin{figure}
\resizebox{!}{0.35\textwidth}
{\includegraphics{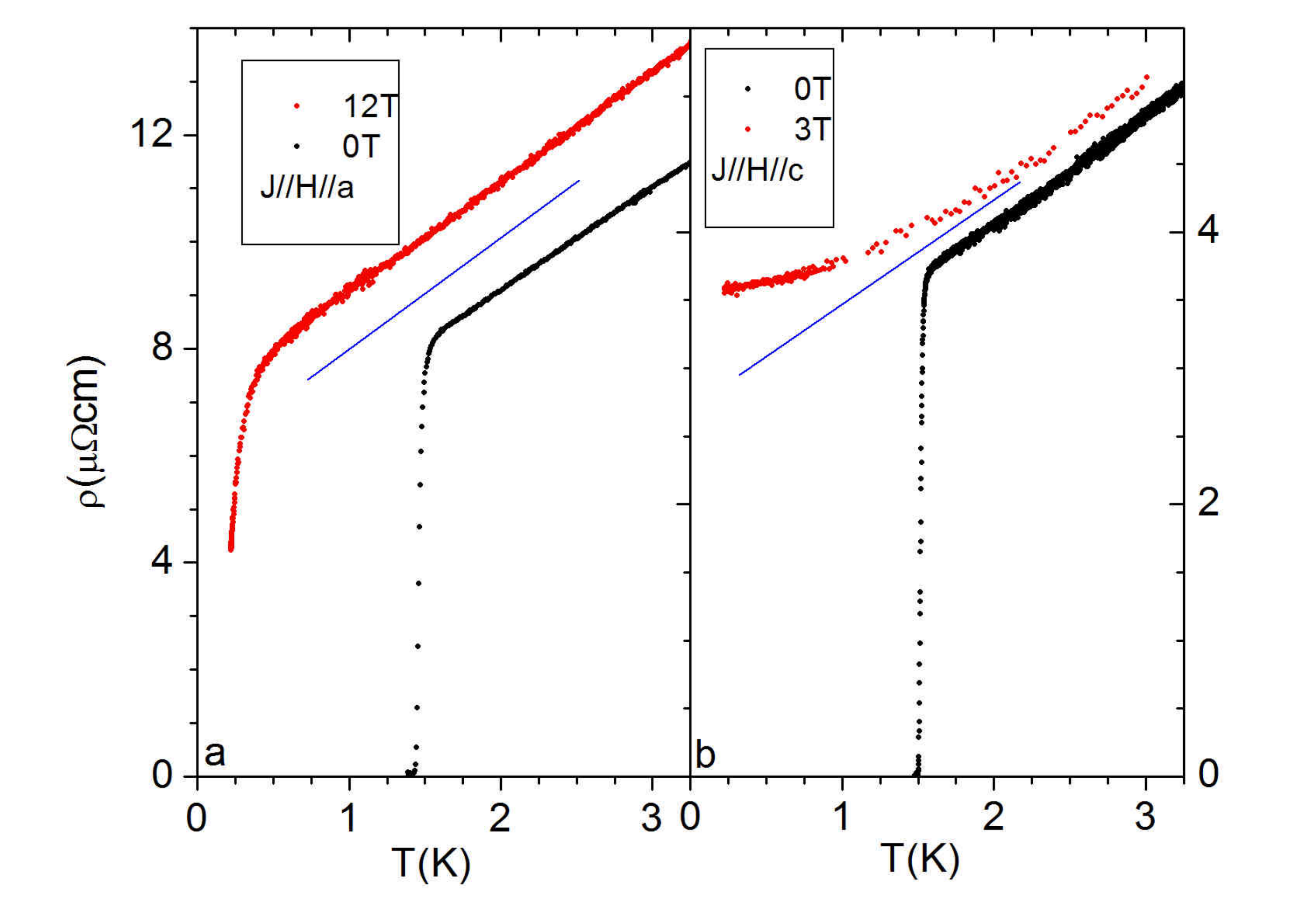}}
\resizebox{!}{0.35\textwidth}
{\includegraphics{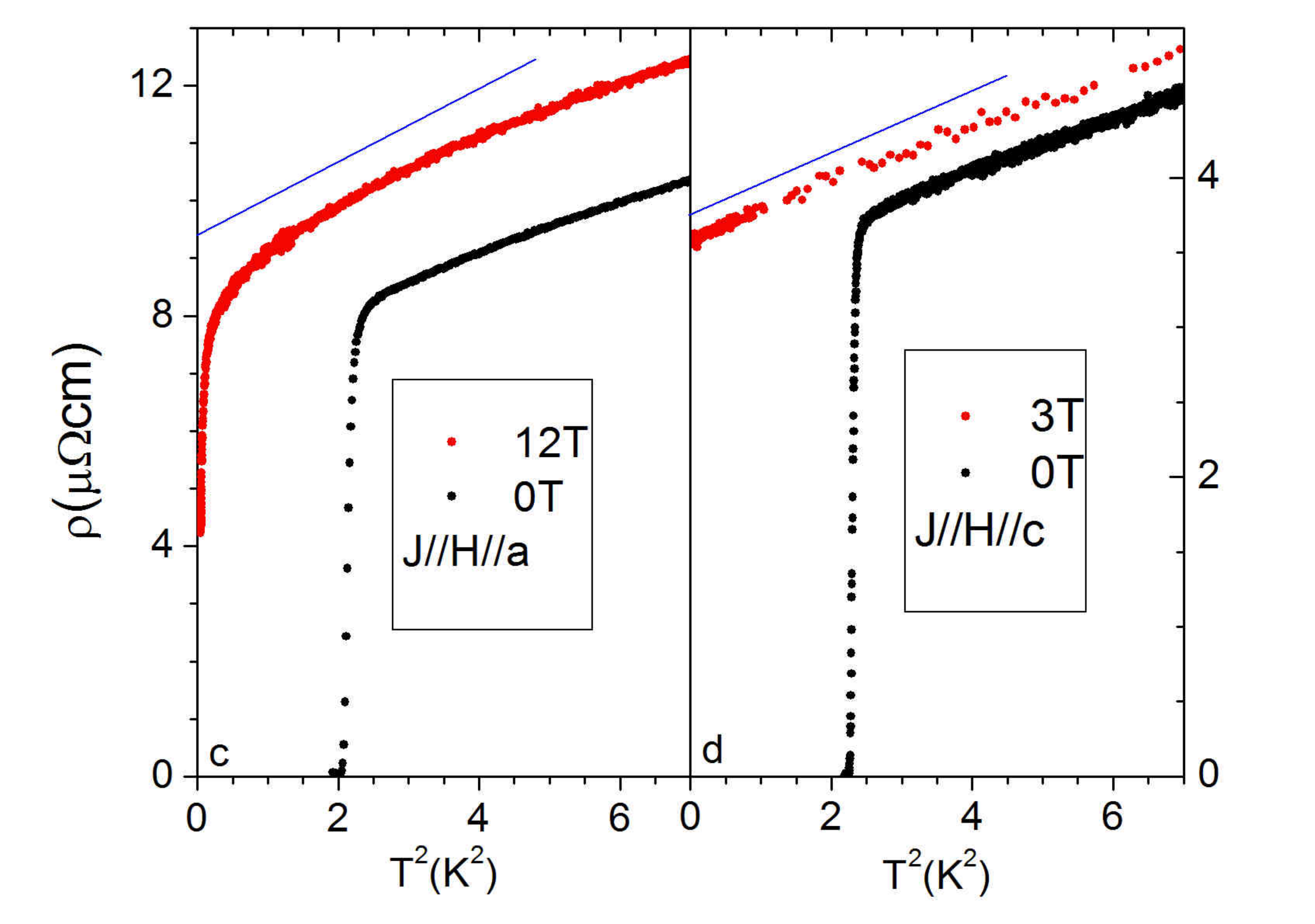}}
\caption{Resistivity vs. temperature for a current along the a-axis (a) and along the c-axis (b). Lower panels present the same data as a function of T$^2$. Solid lines are guides to eye. Resistivity is T-linear along the a-axis and T-square along the b-axis.}
\end{figure}

Single crystals of URu$_{2}$Si$_{2}$ with a residual resistivity ratio (RRR) of 45 were grown by the Czochralski method in a tetra-arc furnace. Two crystals (dimensions:$2\times1\times0.2 mm^{3}$ , each along a principal axis) were used for measurements. The Seebeck coefficient was measured with a standard one-heater-two-thermometer set-up using RuO$_{2}$ thermometers.

Fig. 1a presents the superconducting upper critical field, H$_{c2}(T)$, determined by resistivity measurements. Here H$_{c2}(T)$ is the field at which the resistivity of the sample becomes half of its normal-state value at a given temperature. We checked that changing this 50 percent criterion to 90 or 10 percent does not affect the profile of  H$_{c2}(T)$. The anisotropy found here is very similar to what was previously reported and analyzed by Brison \emph{et al.}\cite{brison}. In particular, as seen in the inset, the slope of H$_{c2}(T)$ near T$_c$, which is governed by the orbital limit,  was found to be three times larger when the field was oriented along the a-axis (See table I). As $\frac{dH_{c2}}{dT}|_{T_{c}}$ is set by the magnitude of the superconducting coherence length, $\xi$, its anisotropy directly yields the anisotropy of $\xi$. Now, in a BCS superconductor, the coherence length at T=0 is set by the Fermi velocity: $\xi_0=\frac{0.18\hbar v_{F}}{k_{B}T_{c}}$. This would imply that the Fermi velocity is \emph{lower} along the c-axis than along the a-axis.

The temperature dependence of the resistivity of the two samples is illustrated in Fig. 1b. As found long ago\cite{palstra2}, and frequently confirmed afterwards, charge conductivity is more than twice larger when the current is along the c-axis at room temperature. This anisotropy presents a modest and continuous increase with cooling down to the onset of the hidden-order transition. The phase transition leads to a sudden drop in this anisotropy which increase steadily afterwards to attain 2.2 at the onset of superconducting transition. Now, in the Drude-Boltzmann picture, the ratio of conductivities is directly related to the ratio of Fermi velocities $\frac{\sigma_c}{\sigma_a}=(\frac{v^{c}_{F}}{v^{a}_{F}})^2$.   This would imply that the Fermi velocity is \emph{larger} along the c-axis than along the a-axis. Thus, we are in presence of a a paradox. As sketched in the inset of each panel, the topology of the Fermi surface extracted from the two methods of analysis are at odd with each other.

This anomalous discrepancy is highlighted in Table I, which compares the case of URu$_2$Si$_2$ with another well-documented heavy-Fermion superconductor UPt$_3$\cite{joynt}. Charge conductivity and superconducting coherence length extracted from the slope of H$_{c2}$ near T$_{c}$ are both anisotropic. However,  there is an excellent compatibility between $\frac{v^{c}_{F}}{v^{a}_{F}}=1.64$ extracted from $\frac{\xi_c}{\xi_a}$ and $\frac{v^{c}_{F}}{v^{a}_{F}}=1.61$ extracted from $(\frac{\sigma_c}{\sigma_a})^{1/2}$. In  URu$_2$Si$_2$, on the other hand, the ratio of the Fermi velocities obtained by these two alternative methods differ by a factor of four (0.35 vs. 1.48). The intimate link between $\frac{\xi_c}{\xi_a}$, $\frac{\sigma_c}{\sigma_a}$ and $\frac{v^{c}_{F}}{v^{a}_{F}}$ is at the origin of a general rule. In different families of anisotropic superconductors, the upper critical field is larger when the field is oriented along the more conducting orientation. URu$_{2}$Si$_{2}$, in company of NpPd$_5$Al$_2$\cite{aoki2,matsuda}, are the only two superconductors we know to disobey this rule. We will argue below that the absence of a single characteristic energy is at the heart of this apparent discrepancy.

Is the ground state of URu$_{2}$Si$_{2}$ in absence of superconductivity a Fermi liquid? The answer to this question has been hindered by the presence of a large transverse magnetoresistance. In clean samples of this low density and  compensated system, even a modest transverse magnetic field leads to an an upturn in resistivity\cite{kasahara}. In order to probe the system without being overwhelmed by the large transverse magnetoresistance, we measured resistivity of both samples in presence of a magnetic field oriented \emph{parallel} to the current. The results are presented in Fig.2 and reveal a qualitative difference between the two orientations of charge flow. Resistivity along the a-axis, $\rho_a$, is T-linear down to the superconducting transition both at zero field and in presence of a field as strong as 12 T (the largest used in this study), with no visible change in slope. On the other hand, for a current along the c-axis, with the application of a magnetic field strong enough to destroy superconductivity, a T-square behavior emerges as the asymptotic temperature dependence of $\rho_c$ in the T=0 limit. Note that this qualitative difference cannot be detected in the absence of magnetic field.

\begin{figure}
\resizebox{!}{0.35\textwidth}
{\includegraphics{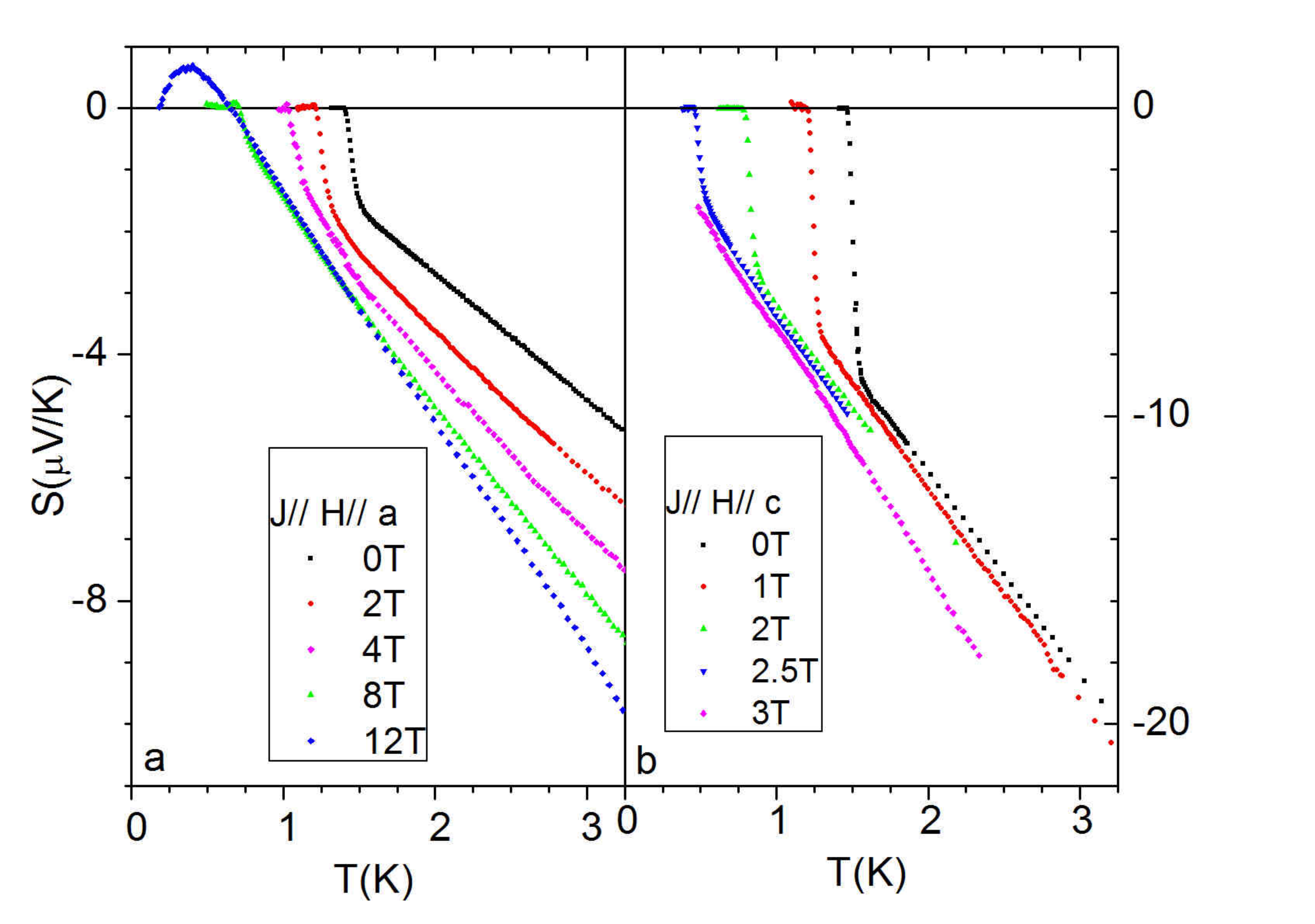}}
\caption{Seebeck coefficient, S, vs. temperature for a thermal current [and gradient] along the a-axis (a) and along the c-axis (b). }
\end{figure}

Fig. 3 presents the thermoelectric data. At zero field, in agreement with previous studies\cite{sakurai,bel}, the Seebeck coefficient is negative and anisotropic. It is roughly twice larger along the c-axis than along the a-axis. With the application of a magnetic field, a qualitative distinction emerges between the two orientations. For a heat current flowing along the c-axis, the Seebeck coefficient, S$_c$ remains negative and  T-linear even in presence of a magnetic field strong enough to destroy superconductivity. On the other hand, in the case of a current flowing along the a-axis, as the applied field is enhanced, the Seebeck coefficient shows a downward deviation from a T-linear dependence and finally it changes sign below 0.8 K, when the magnetic field is strong enough to push T$_{c}$ below this.

The difference between the two orientations is summarized in Fig. 4 which compares resistivity and thermopower.  When the current flows along the c-axis,  Resistivity is T-square and thermopower is T-linear. A finite negative S$_c$/T can be easily estimated. Its magnitude is slightly enhanced in presence of a field, which destroys superconductivity. The situation is drastically different for a current flowing along the a-axis. The Seebeck coefficient changes sign revealing a positive contribution rapidly enhancing with decreasing temperature. Note that superconducting transitions in S(T) and $\rho$ almost coincide each time and the 12 T curves show that the sign change of S is a normal-state property and unrelated to the superconducting transition.

\begin{figure}
\resizebox{!}{0.7\textwidth}
{\includegraphics{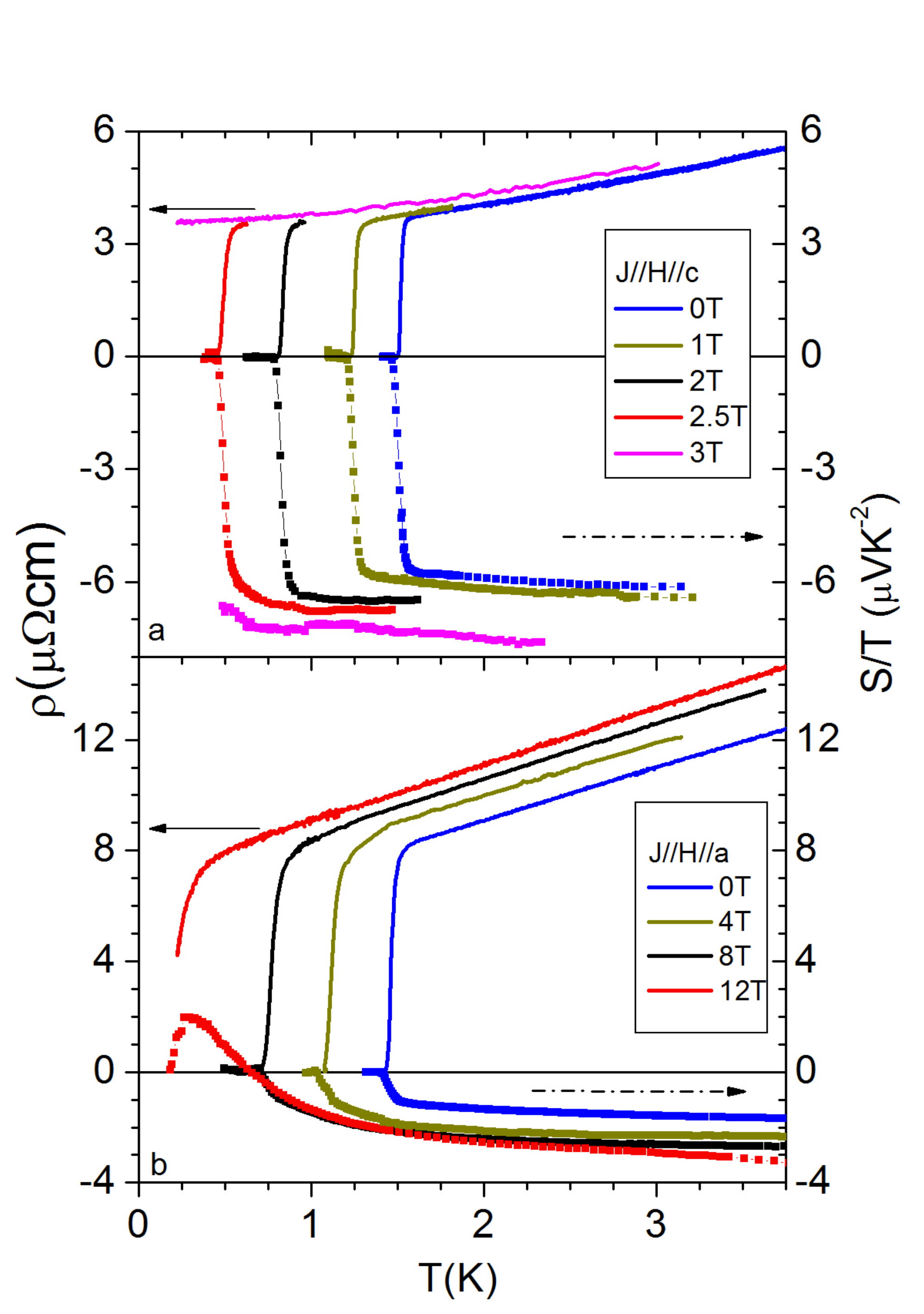}}\caption{Comparison of the temperature dependence of resistivity, $\rho$, and thermopower divided by temperature, $S/T$, as a function of temperature for a current flowing along the c-axis (a) and a current flowing the a-axis (b). In the first case, a T-linear Seebeck coefficient is concomitant with a T-square resistivity.}
\end{figure}

Table I compares the magnitude of the Fermi liquid parameters extracted for the c-axis configuration in URu$_{2}$Si$_{2}$  to those known for UPt$_{3}$\cite{joynt}. The prefactor of T-square resistivity URu$_{2}$Si$_{2}$ obtained here, combined with the reported value of the T-linear specific heat, $\gamma$\cite{maple} yields a Kadowaki-Woods (KW) ratio which is 5.2 times larger than the ``universal'' value (a$_0$=10 $\mu \Omega$ cm mol$^{2}$ K$^{2}$ J$^{-2}$. ) and 17 times larger than the KW ratio along c-axis in UPt$_{3}$. This is not surprising as the KW ratio is expected to increase as the carrier density is reduced\cite{kontani,hussey}. The T-linear Seebeck coefficient for the c-axis configuration in URu$_{2}$Si$_{2}$ is more than twice larger than what was found in the case of UPt$_{3}$\cite{jaccard}. In other words, while the electronic entropy \emph{per volume} (measured by the size of $\gamma$) is 7 times larger in UPt$_{3}$, the entropy \emph{per carrier} is 2.5 times larger in URu$_{2}$Si$_{2}$. The dimensionless ratio of thermopower to electronic specific heat,  $q=\frac{SN_{AV}e}{T\gamma}$  is 11, an order of magnitude larger than what is commonly found in various metals with a half-filled band\cite{behnia2} and 18 times than in UPt$_{3}$, another consequence of a small carrier density.

\begin{table}
\begin{ruledtabular}
\begin{tabular}{|c|c|c|c|c|}
&\multicolumn{2}{c|} {$UPt_{3}$}&
\multicolumn{2}{c|}{$URu_{2}Si_{2}$}\\

\hline
{} & a-axis & c-axis & a-axis & c-axis\\
\hline
$dH_{c2}/dT$ (T K$^{-1}$) & -4.4 & -7.2 & -11.5 & -4.1 \\
\hline
$\xi_{c}$/$\xi_{a}$& \multicolumn{2}{c|}{1.64} & \multicolumn{2}{c|}{0.35}\\
\hline
$\sigma_{c}$/$\sigma_{a}$ & \multicolumn{2}{c|}{2.6} & \multicolumn{2}{c|}{2.2}\\
\hline
$\gamma$ (mJ mol$^{-1}$K$^{-2}$)& \multicolumn{2}{c|}{440} & \multicolumn{2}{c|}{65}\\
\hline
A ($\mu\Omega cm K^{-2}$)& 1.5 & 0.55 & $-$ & 0.22\\
\hline
A/$\gamma^{2}$ (a$_0$) & 0.7 & 0.3 & $-$ & 5.2 \\
\hline
S/T ($\mu$ V K$^{-2}$) & 2.5 & 2.5 & $-$ & -7$\pm$1 \\
\hline
q & 0.6 & 0.6 & $-$ & -11$\pm$1 \\
\end{tabular}
\end{ruledtabular}
\caption{\label{table1} A comparison of physical parameters and their anisotropy in UPt$_{3}$ and in URu$_{2}$Si$_{2}$. }
\end{table}

The in-plane configuration is more enigmatic and remains beyond a simple Fermi-liquid picture. As seen in Fig. 4, below 4 K, $\rho_{a}$ is well described by $\rho_{a}=\rho_{0} + BT$ at all fields and the magnitude of T-linear coefficient (B=1.8 $\mu\Omega$ cm K$^{-1}$) shows little change from zero to to 12 T. Note that this does not necessarily imply that the T-linear behavior would survive down to zero temperature when the the magnetic field exceeds H$_{c2}$. An eventual emerging T$^{2}$ behavior, however, would be restricted to a narrow temperature window and associated with a very large prefactor, pointing to a small characteristic energy scale, much smaller than the one associated with the c-axis. This singular case of anisotropic electron-electron scattering is to be linked with the anomalous anisotropy of the superconducting coherence length and calls for a deep theoretical attention. According to our results, the Cooper pairs extend over a longer distance in the basal plane, which conducts less but is host to the enhanced scattering among electrons.

A link between pairing and T-linear resistivity has been detected in several families of unconventional superconductors\cite{doiron}.  In URu$_{2}$Si$_{2}$, future transport measurements under pressure and magnetic field would allow to determine the profile of the inelastic scattering beyond the critical pressure destroying both the hidden-order and superconductivity\cite{hassinger,hassinger2}. In cuprate superconductors, resistivity becomes purely T$^2$, when superconductivity is destroyed by overdoping\cite{nakamae}. But, at lower doping levels, the resistivity of the ground state recovered by field-induced destruction of superconductivity always contains a T-linear component\cite{cooper}. CeRhIn$_{5}$ is another relevant case. While a Fermi-liquid behavior emerges at low enough temperatures, the pressure-induced destruction of superconductivity is also accompanied by a drastic reduction in the electron-electron scattering\cite{knebel}.

The temperature-dependence of the Seebeck coefficient is more complex.  The sign change suggests the presence of a positive and continuously increasing $S/T$ superposed on a constant negative one. Theoretically, in presence of quantum criticality, $S/T$ is expected to follow $\gamma$ and logarithmically diverge \cite{paul}, but what has been found in the case of CeCoIn$_{5}$\cite{izawa} is not as simple. Qualitatively, the low-temperature departure from a T-linear thermopower confirms a reduced in-plane energy scale. Experiments at higher fields and lower temperatures are required, however, to pin down the asymptotic zero-temperature behavior of the Seebeck coefficient as well as its magnitude and sign for the in-plane configuration.

In summary, our transport measurements show that the electronic properties in URu$_{2}$Si$_{2}$ lack a single Fermi-liquid energy scale for all directions. This  may be the key to explain the anomalous anisotropy of the superconducting coherence length.

This work is part of DELICE, a project supported by the Agence Nationale de Recherche(ANR-08-BLAN-0121-02). Z.Z. acknowledges a scholarship granted by China Scholarship Council.

\end{document}